\documentclass[twocolumn,showpacs,preprintnumbers,amsmath,amssymb]{revtex4-1}
\usepackage{amsfonts, amsmath, bm, bbm, graphicx, epsfig, epstopdf}
\usepackage{dcolumn}
\usepackage{textcomp}

\usepackage{soul}

\begin{document}

\bibliographystyle{apsrev}

\title{Large Energy Superpositions via Rydberg Dressing}

\author{Mohammadsadegh~Khazali, Hon Wai~Lau, Adam~Humeniuk and Christoph Simon}
\affiliation{Institute for Quantum Science and Technology and Department of Physics
and Astronomy, University of Calgary, Calgary T2N 1N4, Alberta, Canada}
\date{\today}
\begin{abstract}
We propose to create superposition states of over 100 Strontium atoms being in a ground state or metastable optical clock state, using the Kerr-type interaction due to Rydberg state dressing in an optical lattice. The two components of the superposition can differ by of order  300 eV in energy, allowing tests of energy decoherence models with greatly improved sensitivity. We take into account the effects of higher-order nonlinearities, spatial inhomogeneity of the interaction, decay from the Rydberg state, collective many-body decoherence, atomic motion, molecular formation and diminishing Rydberg level separation for increasing principal number.
\end{abstract}

\maketitle

\section{Introduction}
\label{sec:introduction}
There are currently many efforts towards demonstrating fundamental quantum effects such as superposition and entanglement in macroscopic systems \cite{Monroe,Brune,Agarwal,Arndt,Sorensen,Friedman,Julsgaard,Esteve,Riedel,OConnell,Lvovsky,Bruno,Palomaki,Vlastakis,
Arndt-Hornberger,Hon-Wai}. One relevant class of quantum states are so-called cat states, i.e. superposition states involving two components that are very different in some  physical observable, such as position, phase or spin. Here we propose a method for creating such large superpositions in energy. This is relevant in the context of testing proposed quantum-gravity related energy decoherence  \cite{Milburn,Gambini,Blencowe}.

Our method relies on the uniform Kerr-type interaction that can be generated between atoms by weak dressing with a Rydberg state \cite{Johnson,Henkel-Thesis,Pfau-Rydberg}. This can be used to generate cat states similarly to the optical proposal of Ref. \cite{Yurke-Stoler}. Using an optical clock state in Strontium as one of the two atomic basis states makes it possible to create large and long-lived energy superposition states. The superposition can be verified by observing a characteristic revival. We analyze the effects of relevant imperfections including higher-order nonlinearities, spatial inhomogeneity of the interaction, decay from the Rydberg state, atomic motion in the optical lattice, collective many-body decoherence triggered by black-body induced transitions, molecular formation, and diminishing Rydberg level separation for increasing principal number. Our scheme significantly improves the precision of energy decoherence detection.


\begin{figure}[h]
\scalebox{0.50}{\includegraphics*[viewport=50 75 510 485]{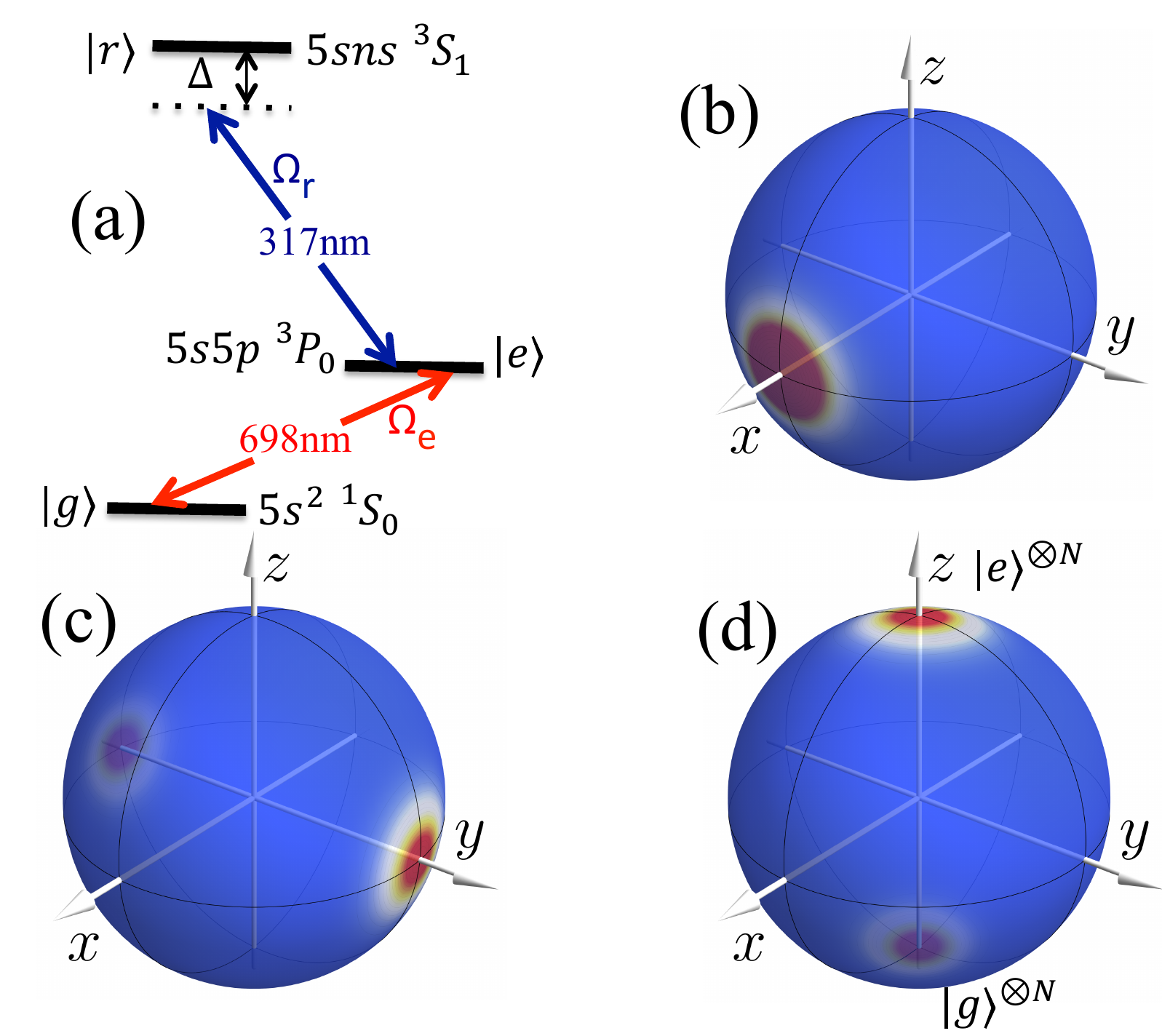}}
\centering
\caption{ (Color online) Proposed scheme for creation of large energy superposition. (a) Level scheme in Strontium. The pseudo-spin states are the singlet ground state $|g\rangle$ and a long lived excited triplet state $|e\rangle$. An off-resonant laser field $(\Omega_{r})$ dresses  the excited state with the Rydberg level $|r\rangle$. This creates a Kerr-type interaction between the atoms in the excited state. The resonant laser field $(\Omega_{e})$ is applied for population rotation. (b-d) The evolution of the Husimi distribution of the collective spin state on the Bloch sphere. Application of the Kerr-type interaction splits the initial coherent spin state (CSS) (b) into a superposition of two CSS at opposite poles of the Bloch sphere (c).  Applying a $\pi /2$ rotation along the $x$ axis following the cat creation process results in a superposition of all atoms being in the ground or excited state. } \label{Scheme}
\end{figure}

Previous related, but distinct, work includes Ref. \cite{Simon-Jaksch} who briefly discussed the creation of energy superposition states in Strontium Bose-Einstein condensates based on collisional interactions.  Ref. \cite{Ghobadi} proposed the creation of energy superposition states of light, and ref. \cite{ion-Cat}  reported the realization of 14-ion GHZ state, with 24 eV energy separation, but without mentioning the energy superposition aspect. The present proposal promises much greater sensitivity to energy decoherence thanks to a much longer lifetime (compared to Ref. \cite{Ghobadi}) and to both increased size and longer lifetime (compared to Ref. \cite{ion-Cat}). Related work involving Rydberg states includes Refs. \cite{Saffman-Molmer,Opatrny-Molmer}, who performed detailed studies of the creation of moderate-size cat states using Rydberg blockade. The number of atoms is limited to of order ten in these schemes due to competing requirements for the presence and absence of blockade between different Rydberg transitions in the same ensemble. They also don't use metastable optical clock states, resulting in only small differences in energy between the two components. Ref. \cite{Mukherjee} briefly discussed the creation of moderate-size (15 atoms) GHZ type states in Strontium atom chains, without mentioning the energy superposition aspect. Ref. \cite{Mukherjee} uses attractive Rydberg interactions, but not the uniform Kerr-type interaction used in the present work. The number of atoms in Ref. \cite{Mukherjee} is limited by unwanted transitions to other nearby many-body states \cite{Mukherjee-Thesis}.

The paper is organized as follows. We begin with a description of our scheme in Sec.~\ref{sec:Scheme}. In Sec.~\ref{sec:Imperfection} and  \ref{Decoherence} we quantify the effects of the main imperfections and decoherence sources on the fidelity of final cat state. In Sec.~\ref{Cat Size} we find an estimate for size of cat states that can be realized with high fidelity. We then show that our scheme is experimentally realizable in Sec.~\ref{Realization}, followed by a detailed discussion in Sec.~\ref{More imperfection}, demonstrating that the effects of atomic motion, molecular formation, collective many-body decoherence, level mixing and BBR radiation induced decoherence can be suppressed. We conclude the paper in Sec.~\ref{Energy-decoherence} with a discussion of the application of energy superposition states for the detection of energy decoherence.

\section{Scheme}
\label{sec:Scheme}

We now describe our proposal in more detail. In an ensemble of $N$ ultra-cold Strontium atoms trapped in a 3D optical lattice \cite{3D Lattice}, one can consider a two-level system consisting of the singlet ground state $|g\rangle$ and a long-lived excited triplet state $|e\rangle$, which are separated in energy by $1.8$ eV. An interaction between the atoms can be induced by dressing the clock state with a strongly interacting Rydberg level \cite{Johnson,Henkel-Thesis,Pfau-Rydberg} as shown in the level scheme of Fig.~\ref{Scheme}. This induces a light shift (LS) on the atoms which depends on the Rydberg blockade.

\subsection{Kerr-type Rydberg Dressed Interaction}
\label{Ker-type}
When the entire ensemble is inside the blockade radius, the dressing laser couples the state with no Rydberg excitation $|\psi_1\rangle=\otimes_{i}\left|\phi_{i}\right\rangle $ (where $\phi\in\{e,g\}$) to a state where only one of the atoms in the $|e\rangle$ level gets excited to the Rydberg level $|\psi_2\rangle=\sum_{i}\left|\phi_{1}...r_{i}...\phi_{N}\right\rangle$  with an enhanced Rabi frequency $\sqrt{N_{e}}\Omega_{r}$ \cite{kuzmich}, where $N_e$ is the number of atoms in the excited state.
Over the Rydberg dressing process, the Hamiltonian can be diagonalized instantaneously
\begin{equation}
D \equiv UHU^\dagger=\left(\begin{array}{cc}E_{-} & 0\\0 & E_{+}\end{array}\right),
\end{equation}
where $E_{\pm}=\frac{\Delta}{2}(1 \pm \sqrt{1+\frac{N_{e}\Omega_{r}^{2}}{\Delta^2}})$ and
\begin{equation}
U=\left(\begin{array}{cc}\cos(\theta/2) & -\sin(\theta/2)\\ \sin(\theta/2) & \cos(\theta/2)\end{array}\right)
\end{equation}
with $\theta=\tan^{-1}(\frac{\sqrt{N_e}\Omega_r}{\Delta}) $. The Schrodinger equation expressed in the dressed state basis $|\varphi>=U|\psi>$ is
\begin{equation}
i\frac{\partial}{\partial t} \left(\begin{array}{cc} |\varphi_{-} \rangle  \\ |\varphi_{+}\rangle \end{array}\right)=\left(\begin{array}{cc}E_{-} & -i\dot{\theta}/2 \\ i\dot{\theta}/2 & E_{+}\end{array}\right) \left(\begin{array}{cc} |\varphi_{-} \rangle  \\ |\varphi_{+}\rangle \end{array}\right).
\end{equation}

To avoid the scattering of population from the ground dressed state to the excited dressed state, the coupling term $\dot{\theta}=\frac{\sqrt{N_e}\Omega_r \dot{\Delta} -\sqrt{N_e} \Delta \dot{\Omega}_r }{N_e \Omega_r^2+\Delta^2}$ should be smaller than $E_+$ (see realization section \ref{Realization} for examples).

Focusing on the ground dressed state, the effective light shift of the system  is
\begin{equation}
E_-=\frac{\Delta}{2}(1 - \sqrt{1+\frac{N_{e}\Omega_{r}^{2}}{\Delta^2}}).
\label{light-shift Eq}
\end{equation}
Within the weak dressing regime $(\frac{\sqrt{N_e}\Omega_r}{\Delta}\ll 1)$ one can Taylor expand the light shift to
\begin{equation}
E_-=\frac{\Delta}{2}[1 -  (1+\frac{1}{2} \frac{N_e \Omega_r^2}{\Delta^2} - \frac{1}{8} \frac{N_e^2 \Omega_r^4}{\Delta^4} +O(\frac{N_e \Omega_r^2}{\Delta^2})^3 ) ],\label{Expansion Eq}
\end{equation}
which can be simplified to $E_- \approx (N_e^2-\frac{N_e}{w^2})\frac{\chi_0}{2}$, with $w=\frac{\Omega_r}{2\Delta}$ and $\chi_0=2w^4\Delta$. Therefore adiabatic weak dressing of atoms to the Rydberg level imposes an effective Kerr-type Hamiltonian
\begin{equation}
H= (\hat{N_e}^2-\frac{\hat{N_e}}{w^2})\frac{\chi_0}{2}\label{Kerr-type-Eq}
\end{equation}
on the atoms within the blockade radius. The effects of higher order terms in the Taylor expansion are discussed in Sec.~\ref{Higher Order Non-linearities} and Fig.~\ref{wVsN}.

\subsection{Generation of Cat State on the Equator of the Bloch Sphere}

 The two levels $|g_{i}\rangle$ and $|e_{i}\rangle$ for each atom are equivalent to a spin $1/2$ system with Pauli matrices $\sigma_{x}^{(i)}=(|g_i\rangle\langle e_i|+|e_i\rangle \langle g_i|)/2$, $\sigma_{y}^{(i)}=i(|g_i\rangle\langle e_i|-|e_i\rangle \langle g_i|)/2$ and $\sigma_{z}^{(i)}=(|e_i\rangle\langle e_i|-|g_i\rangle \langle g_i|)/2$ acting on the atom at site $i$. We define collective spin operators $S_{l}=\sum_{i=1}^{N}\sigma_{l}^{(i)}$. A coherent spin state (CSS) is defined as a direct product of single spin states \cite{CSS}
\begin{equation}
|\theta,\phi \rangle=\otimes _{i=1}^{N}[\cos{\theta}|g \rangle_{i}+\sin{\theta} e^{i \phi} |e \rangle_{i}],
\end{equation}
where all the spins are pointing in the same direction, and $\phi$ and $\theta$ are the angles on the (collective) Bloch sphere. The CSS can also be represented as \cite{CSS}
\begin{equation}
|\eta \rangle= |\theta,\phi \rangle=(1+|\eta|^{2})^{-N/2} \sum_{N_{e}=0} ^{N} \eta^{N_{e}} \sqrt{C(N,N_e)}|N;N_{e}\rangle,
\end{equation}
where $\eta=\tan(\theta/2)e^{-i\phi}$,
$C(N,N_e)\equiv\left(\begin{array}{c} N\\ N_{e} \end{array}\right)$ and  $|N;N_{e} \rangle=\frac{1}{\sqrt{C(N,N_e)}}\sum_{i1<i2<...<iN_{e}}^{N}|g_{1}...e_{i1}...e_{iN_{e}}...g_{N}\rangle$ is the Dicke state of $N_{e}$ excited atoms, where $|N;N_{e} \rangle$ is an alternative representation of the $|J \, M\rangle$ basis with $N=2J$ and $N_e=J+M$.

Let us now discuss the time evolution of an initial CSS $|\eta\rangle$ under the Kerr-type interaction of Eq.~(\ref{Kerr-type-Eq}). The state evolves as
\begin{equation}
|\psi(t) \rangle= (1+|\eta|^{2})^{-N/2} \sum_{N_{e}=0} ^{N} \eta^{N_{e}} e^{-i H t} \sqrt{C(N,N_e)}|N;N_{e}\rangle.
\end{equation}
At the ``cat creation'' time $\tau_{c}=\frac{\pi}{ \chi_{0}}$ the linear term of Eq.~\ref{Kerr-type-Eq} creates a phase rotation, which changes the state to $|\eta' \rangle=|e^{-i\frac{N_e \chi_0}{2w^2}\tau_c}\eta \rangle$. The quadratic term produces coefficients of $(1)$ and $(-i)$ for even and odd $N_{e}$'s respectively. The state can then be rewritten as a superposition of two CSS, namely
\begin{equation}
|\psi(\tau_{c}) \rangle=\frac{1}{\sqrt{2}} (e^{i \frac{\pi}{4}}|\eta' \rangle+e^{-i \frac{\pi}{4}}|-\eta' \rangle)
\end{equation}
in analogy with Ref. \cite{Yurke-Stoler}. Continuing the interaction for another $\tau_{c}$, one can observe the revival of the initial CSS. This revival can be used as proof for the successful creation of a quantum superposition at $\tau_c$, since a statistical mixture of CSS at $\tau_c$ would evolve into another mixture of separate peaks \cite{Hon-Wai,Dalvit}.

\subsection{Creating the Energy Cat}
\label{Steps Towards the Energy Cat}

To create an energy superposition state we thus have to apply the following steps. Starting from the collective ground state $|g \rangle ^{\otimes N}$, we apply a $\pi/2$ pulse on the $|e\rangle-|g\rangle$ transition that results in the maximum eigenstate of the $S_{x}$ operator $|\eta=1 \rangle=(\frac{|e \rangle  +| g \rangle}{\sqrt{2}})^{\otimes N}$, as shown in Fig.~\ref{Scheme}(b). Since the atoms are confined to the ground states of optical lattice traps, the position-dependent phase factors associated with laser excitation of the clock state are constant over the course of the experiment and can be absorbed into the definition of the atomic basis states (detailed discussion can be found in Sec.~\ref{atomic motion}). We now apply the Kerr-type interaction. The large coefficient of the linear term in the Hamiltonian leads to a rotation of the created cat state on the equator of Bloch sphere. Applying accurate interaction timing, the state can be chosen to be a superposition of two CSS pointing to opposite directions along the $y$ axis on the Bloch sphere
$| \psi(\tau_{c})\rangle=\frac{1}{\sqrt{2}}(e^{i \frac{\pi}{4}}   |\eta=i \rangle+ e^{-i \frac{\pi}{4}} |\eta=-i\rangle)$,
see Fig.~\ref{Scheme}(c) and inset (a) of Fig.~\ref{wVsN}. For example, a timing precision of $\delta \tau_c=\frac{2w^2}{5 \pi\sqrt{N}}\tau_c$ results in an adequate phase uncertainty of $\delta\phi=\frac{1}{5 \sqrt{N}}$ (examples can be found in the realization Sec.~\ref{Realization}). Applying another $\frac{\pi}{2}$ pulse on the created cat state results in $\frac{|e \rangle ^{\otimes N} +| g \rangle^{\otimes N}}{\sqrt{2}}$, which is a superposition of all the atoms being in the ground and excited states, as shown in Fig.~\ref{Scheme}(d). The created state is a superposition of two components with very different energies. To verify the creation of the energy cat state one needs to rotate the state back to the equator and detect the revival of the initial CSS under the Kerr-type interaction, see also the inset of Fig.~\ref{wVsN}(b).

\section{Imperfections}
\label{sec:Imperfection}

In this section we quantify the effects of the most important imperfections with direct impact on the achievable cat size. Other sources of imperfections, which can be made to have relatively benign effects on our scheme, are discussed in Sec.~\ref{More imperfection}. 

\subsection{Higher Order Non-linearities}
\label{Higher Order Non-linearities}

\begin{figure}
\centering
\scalebox{0.43}{\includegraphics*[viewport=30 210 580 500]{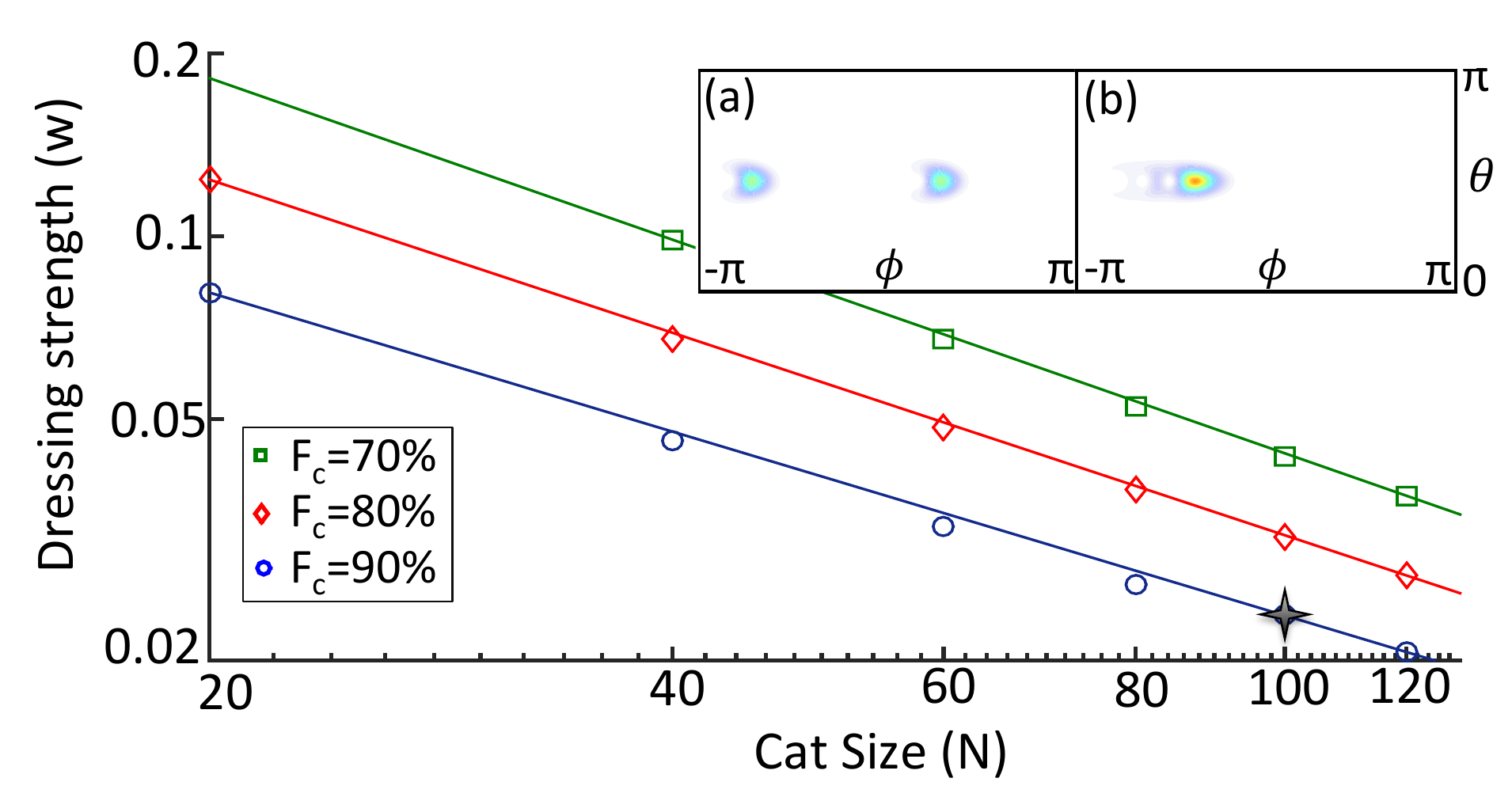}}
\caption{ (Color online) Effect of higher than second order nonlinearities (from the higher orders of Eq.~\ref{Expansion Eq}) on the fidelity of the cat state. The weak dressing parameter $(w=\frac{\Omega_{r}}{2 \Delta})$ has to be reduced for larger atom numbers $N$ in order to keep a fixed fidelity $F_{nl}$ ($F_{nl}=0.7$ (green), $0.8$ (red), $0.9$ (blue) from top to bottom). The inset shows the Husimi Q function for an $N=100$ cat state (a) with $F_{nl}=0.9$ (corresponding to the black cross in the main figure), as well as the corresponding revival (b). The approximate revival of the initial CSS at the time $t=2 \tau_c$ proves the existence of a quantum superposition at $t=\tau_c$.} \label{wVsN}
\end{figure}
First, we only considered the linear and quadratic terms in $N_e$ in our Hamiltonian, which is accurate for very weak dressing. Applying stronger dressing fields yields a stronger interaction, but also increases the importance of higher order terms in Eq.~(\ref{Expansion Eq}). To quantify the effects of these higher orders, we calculate the fidelity of the cat state $(|\psi'(\tau_{c}) \rangle)$ generated based on Eq.~(\ref{light-shift Eq}) with respect to the closest ideal cat state,
\begin{equation}
F_{nl}=\mbox{max}_{\theta,\phi,\alpha, \tau_c}|\langle \psi'(\tau_{c}) | \frac{1}{\sqrt{2}}(|\theta,\phi \rangle+e^{i \alpha}|\pi-\theta,\phi+\pi \rangle)|^{2}.
\end{equation}
Fig.~\ref{wVsN} shows that the weak dressing parameter $w=\frac{\Omega_r}{2\Delta}$ has to be reduced for larger atom numbers in order to achieve a desired fidelity.

\subsection{Effects of Interaction Inhomogeneities}
\label{sec:Inhomogeneities}

\begin{figure}
\centering
\scalebox{0.34}{\includegraphics*[viewport=0 80 800 420]{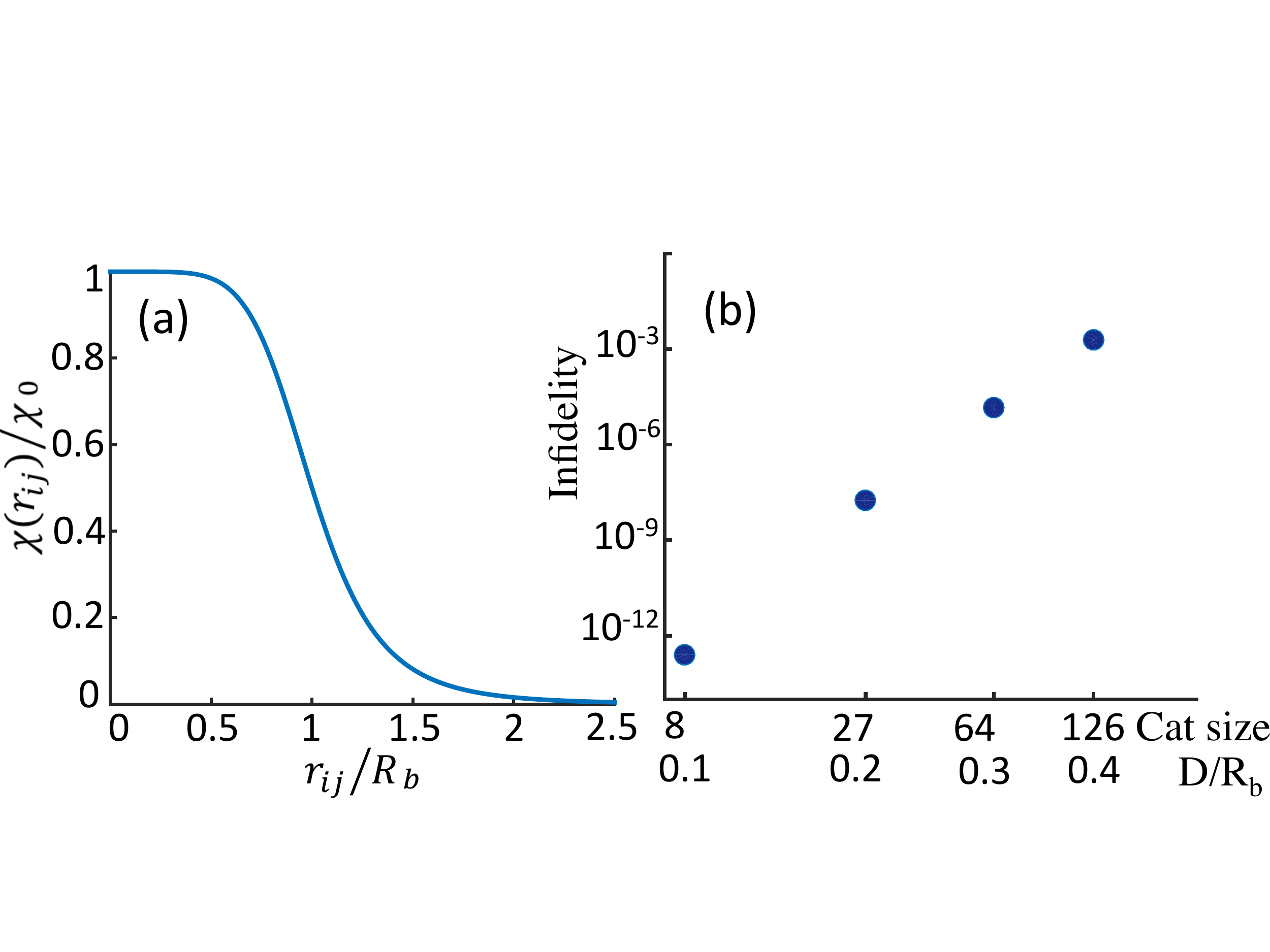}}
\caption{ (Color online) Effect of interaction inhomogeneity. (a) Plateau-type interaction between each pair of atoms dressed to the Rydberg state. The interaction is uniform for separations up to of order the blockade radius. (b) Infidelity caused by interaction inhomogeneity as a function of cat size $(N)$, for a constant blockade radius. Non-linear fidelity is set to  $F_{nl}=0.9$, the blockade radius $R_{b}=3.6\mu$m is created by Rydberg dressing to $n=80$, and the atoms are considered to be in a cubic trap with space diagonal $D$ and lattice spacing of $200$nm.} \label{Inhomogeneity Fig}
\end{figure}

We also considered a uniform blockade over the entire medium, leading to a homogeneous interaction. In practice the interaction is not perfectly homogeneous. One can apply fourth order perturbation theory to find the interaction of the entire weakly dressed system as \cite{Pohl1,Pohl2}
\begin{equation}
 \hat{H}=\sum_{i<j} \chi(r_{ij}) \hat{\sigma}^{i}_{ee} \hat{\sigma}^{j}_{ee}-\frac{\Omega^2}{4\Delta}\hat{N}_e.
\end{equation}
The many-body interaction is the sum of binary interactions
\begin{equation}
\chi(r_{ij})=\chi_{0} \frac{R_{b}^{6}}{r_{ij}^{6}+R_{b}^{6}},
\end{equation}
where $R_{b}=|\frac{C_{6}}{2\Delta}|^{1/6}$ is the blockade radius in the weak dressing regime. This binary interaction has a plateau type nature, see Fig.~\ref{Inhomogeneity Fig}(a). The inhomogeneity of the interaction introduces a coupling to non-symmetric states, since the Hamiltonian no longer commutes with the total spin operator ($[S^2,H]\neq 0$). We evaluate the fidelity of a cat state created by the realistic non-uniform interaction with respect to the ideal cat state. Writing the pair interactions $\chi(r_{ij})$ in terms of small fluctuations $\epsilon_{ij}$ around a mean value $\chi_{m}$,
we decompose the Hamiltonian into a sum of two commuting terms, $\hat{V}_{H}=\sum\limits_{i<j} \chi_m \hat{\sigma}^{i}_{ee} \hat{\sigma}^{j}_{ee}-\frac{\Omega^2}{4\Delta}\hat{N}_e=\chi_{m} (\frac{\hat{N}_e^2-\hat{N}_e}{2})-\frac{\chi_0}{2w^2}\hat{N}_e \approx \frac{\chi_{m}}{2}\hat{N}_e^2-\frac{\chi_0}{2w^2}\hat{N}_e$ and $\hat{V}_{IH}=\sum\limits_{i<j} \epsilon_{ij} \hat{\sigma}^{i}_{ee} \hat{\sigma}^{j}_{ee}$, corresponding to the homogeneous and inhomogeneous parts respectively. While the homogeneous part leads to an ideal cat state, the inhomogeneous part reduces the fidelity by a factor $F_{IH}=|\langle \eta=1|e^{-i\hat{V}_{IH}\tau_c}|\eta=1\rangle|^2$, where $|\eta=1 \rangle=(\frac{|e\rangle+|g\rangle}{\sqrt{2}})^N$ is the initial CSS.
Taylor expanding the inhomogeneous part of the evolution operator one obtains an estimate for the fidelity as explained in Appendix \ref{supp}.
Fig.~\ref{Inhomogeneity Fig}(b) shows the resulting infidelity as a function of cat size for constant blockade radius.

\section{Decoherence}
\label{Decoherence}
The main source of decoherence in our system is depopulation of the Rydberg level which also determines the lifetime of the dressed state $(\tau_{\tilde{e}} \approx \tau_r w^{-2})$. In this section we identify different Rydberg decay channels and discuss their effects on the fidelity of the cat state. Loss due to collisions is reduced by the use of an optical lattice trap with a single atom per site. Ref. \cite{Lattice-Lifetime} implemented a Strontium optical clock using a blue-detuned lattice (trap laser wavelength 390 nm) with a collision-limited lifetime of $100$s, demonstrating that loss due to the trap laser can be made negligible. Other sources of decoherence including blackbody radiation induced transitions, collective many-body decoherence and molecular formation will be discussed in Sec.~\ref{More imperfection}.

\subsection{Rydberg Decay Channels}
\label{A coefficient}
The main source of decoherence in our system is depopulation of the Rydberg level which also determines the lifetime of the dressed state $(\tau_{\tilde{e}} \approx \tau_r w^{-2})$. The Rydberg state depopulation rate can be calculated as the sum of spontaneous transition probabilities to the lower states (given by Einstein A-coefficients) \cite{A-coefficient1,A-coefficient2,A-coefficient3}
\begin{equation}\label{Life-time}
\tau_{r}^{-1}=\sum\limits_{f}{A_{if}}=\frac{2e^{2}}{3\epsilon_{0}c^{3}h} \sum\limits_{E_f<E_i} \omega_{if}^{3}\, |\langle i|\vec{r}|f\rangle|^{2},
\end{equation}
where $\omega_{if}=\frac{E_f-E_i}{\hbar}$ is the transition frequency
 and $\langle i|\vec{r}|f\rangle$ is the dipole matrix element
between initial and final states (see Appendix~\ref{dipole matrix elements}).
The summation is only over the states $|f\rangle$ with lower energies compared to the initial state.
Using a cryogenic environment \cite{Cryogenic}, black-body radiation induced transitions are negligible, see Sec.~\ref{BBR} for detailed discussion.

Considering the dressing to $5sns \,{}^3S_{1}$ in our proposal, the possible destinations of dipole transitions are limited to $^3P_{0,1,2}$, due to the selection rules.  Around 55\% of the transferred population will be trapped within the long-lived $^3P_2$ states, which we refer to as qubit loss. Around 35\% of the population is transferred to $^3P_1$ states, which mainly decay to the ground state $|g \rangle =5s^2\, ^1S_0$ within a short time
 (e.g. $\tau_{5s5p\, ^3P_1}=23 \mu s$ \cite{3P0lifetime}), which we refer to as de-excitation. The remaining 10\% of the population is transferred to $^3P_0$ states. Half of this population ($5\%$ of the total) contributes to qubit loss, bringing the total loss to 60 $\%$, while the other half (also $5\%$ of the total) is transferred to the excited state, which effectively causes dephasing of $|\tilde{e}\rangle$ because the photon that is emitted in the process contains which-path information about the qubit state.

\subsection{Effects of Rydberg Decoherence on the Cat State}

The three decoherence types discussed in the previous sub-section have different effects on the cat state. Loss and de-excitation completely destroy the cat state if they occur, while dephasing is both unlikely and relatively benign. We now explain these statements in more detail.

The majority (60\%) of the dressed state's decay goes to non-qubit states $|\tilde{e} \rangle \Rightarrow \delta |\tilde{e} \rangle |0\rangle_p +\sqrt{1-\delta^2} |l \rangle | 1 \rangle_p $, where $\delta^2=e^{-0.6 \gamma_{\tilde{e}} \tau_c}$ and   $|1 \rangle_p$ represents the emitted photon. In addition to loss, 35\% of the dressed state's decay is de-excitation $|\tilde{e} \rangle \Rightarrow \delta |\tilde{e} \rangle |0\rangle_p +\sqrt{1-\delta^2} |g\rangle | 1 \rangle_p $,  where $\delta^2=e^{-0.35 \gamma_{\tilde{e}} \tau_c}$.

Decay of a single dressed state atom transforms an atomic symmetric Dicke state $|N;N_e \rangle$ into a combination of the original state $|N;N_e \rangle$, a symmetric Dicke state $|N;N_e-1 \rangle$ with one fewer excitation, and $N$ different other Dicke states $(|N-1;N_e-1 \rangle _{\tilde{i}} |l\rangle_i)$ in which the $i$-th atom is transferred to a non-qubit state (the qubit is lost), but which are still symmetric Dicke states for the remaining atoms. The resulting state is $\sqrt{P_0} |N;N_e \rangle |0\rangle_P +  \sqrt{P_{de} N_e} |N;N_e-1 \rangle |1\rangle_P + \sqrt{\frac{P_l N_e}{N}}\sum\limits_{i=1}^N |N-1;N_e-1 \rangle _{\tilde{i}} |l\rangle_i |1\rangle_P $ where $P_{k}=\lambda_k e^{-\lambda_k}$ is the probability of losing/de-exciting $(k=l/de)$ an atom over the cat creation time, with $\lambda_{k}= \gamma_{(k)} \frac{N}{2} \tau_c$ (note that $ N_e \sim \frac{N}{2}$ since the cat creation happens on the equator of Bloch sphere) and $P_0=1-P_l-P_{de}$. Here we focus on the regime where the probability of a single atom decaying is sufficiently small that the probability of two atoms decaying can be ignored.


Tracing over the lost qubit and the photonic state one obtains the density matrix $\rho_c=P_0 \rho_0 + \frac{P_l }{N}\sum\limits_{i=1}^N\rho^i_l+ P_{de} \rho_{de}$, where $\rho_0$ and $\rho_{de}$ are in the symmetric subspace with total spin $(J=\frac{N}{2})$, while the $\rho^i_l$ are in $N$ different symmetric subspaces with total spin $(J=\frac{N-1}{2})$. The $\rho_0$ component corresponds to the ideal cat state. All the other components have very small fidelity with ideal cat states, primarily because the decay happens at a random point in time, which leads to dephasing. For example, de-excitation of an atom  at $(t_{de}\in[0,\tau_c])$, leads to
\begin{eqnarray}\label{expansion}
&&| \psi^{de}_c(t_{de}) \rangle= 2^{-N/2}\sum\limits_{N_{e}=1} ^{N} \sqrt{C(N,N_e)} \\ \nonumber
&& e^{-iE_{(N_e-1)}(\tau_c-t_{de})} \sqrt{N_e}  e^{-iE_{(N_e)}t_{de}}|N;N_{e}-1\rangle,
\end{eqnarray}
where $E_{(N_e-1)}$ represents the dressed state energy of $(N_e-1)$ excited atoms, see Eq.~(\ref{Kerr-type-Eq}).
Inserting the expressions for $E_{N_e}$ and $E_{N_e-1}$, one sees that de-excitation adds a linear term  $(iN_e \chi_0 t_{de})$ to the phase. This creates a rotation around the $z$ axis on the Bloch sphere. The uncertainty in the time of decay $t_{de}$ therefore dephases the cat state, resulting in the formation of a ring on the equator of the Bloch sphere, which has a small overlap with the ideal cat state.  The fidelity of the resulting density matrix compared to an ideal cat state in the same subspace (which corresponds to the case where de-excitation happens at $t_{de}=0$) can be written as $F_{de}=\frac{1}{\tau_{c}} \intop^{\tau_{c}} _{0} |\langle \psi^{de}_c(t_{de})  | \psi^{de}_c(t_{de}=0) \rangle|^2 dt_{de}$. When the size of the cat state is increased from $N=10$ to $N=160$, the fidelity of the generated cat in the de-excited subspace is reduced from $F_{de}=0.2$ to $F_{de}=0.045$. The fidelity in each of the $N$ subspaces where one atom was lost can be calculated in a similar way, yielding equivalent results. The total fidelity in the presence of Rydberg decoherence is then $F_{dc}=P_0+P_lF_l+P_{de}F_{de} \approx P_0$.

About 5\% of Rydberg decoherence will transfer back to the excited state, which acts as dephasing (modeled by a Lindblad operator $|\tilde{e} \rangle \langle \tilde{e}|$). The dephasing operator commutes with the Hamiltonian for cat state creation. Its effect can therefore be studied by having it act on the final cat state. For example, it can cause a sign flip of $|e\rangle$ for the first atom, resulting in a state $(\frac{|e \rangle+i|g \rangle}{\sqrt{2}})(\frac{|e \rangle-i|g \rangle}{\sqrt{2}})^{\otimes(N-1)}+(\frac{|e \rangle-i|g \rangle}{\sqrt{2}})(\frac{|e \rangle+i|g \rangle}{\sqrt{2}})^{\otimes (N-1)}$. Applying the $\pi/2$ rotation results in a new energy cat $\frac{|g\rangle |e\rangle^{N-1}+|e\rangle|g\rangle^{N-1}}{\sqrt2}$, which is clearly still a large superposition in energy. So the effect of dephasing errors is relatively benign.
Moreover, given the small relative rate of dephasing compared to loss and de-excitation,
the probability of having a sign flip over the cat creation time for the case with decoherence fidelity of $F_{dc}=0.8$ (considered in Fig.~\ref{Nvsn}) will only be 1$\%$.

In conclusion, the fidelity of the cat state is, to a good approximation, equal to the probability of not losing or de-exciting any qubits over the cat creation time, $F_{dc} =P_0=e^{-0.95 \frac{N}{2}\gamma_{\tilde{e}}\tau_c}$.

\section{Estimate of Realizable Cat Size}
\label{Cat Size}

\begin{figure}
\centering
\scalebox{0.34}{\includegraphics*[viewport=0 55 680 458]{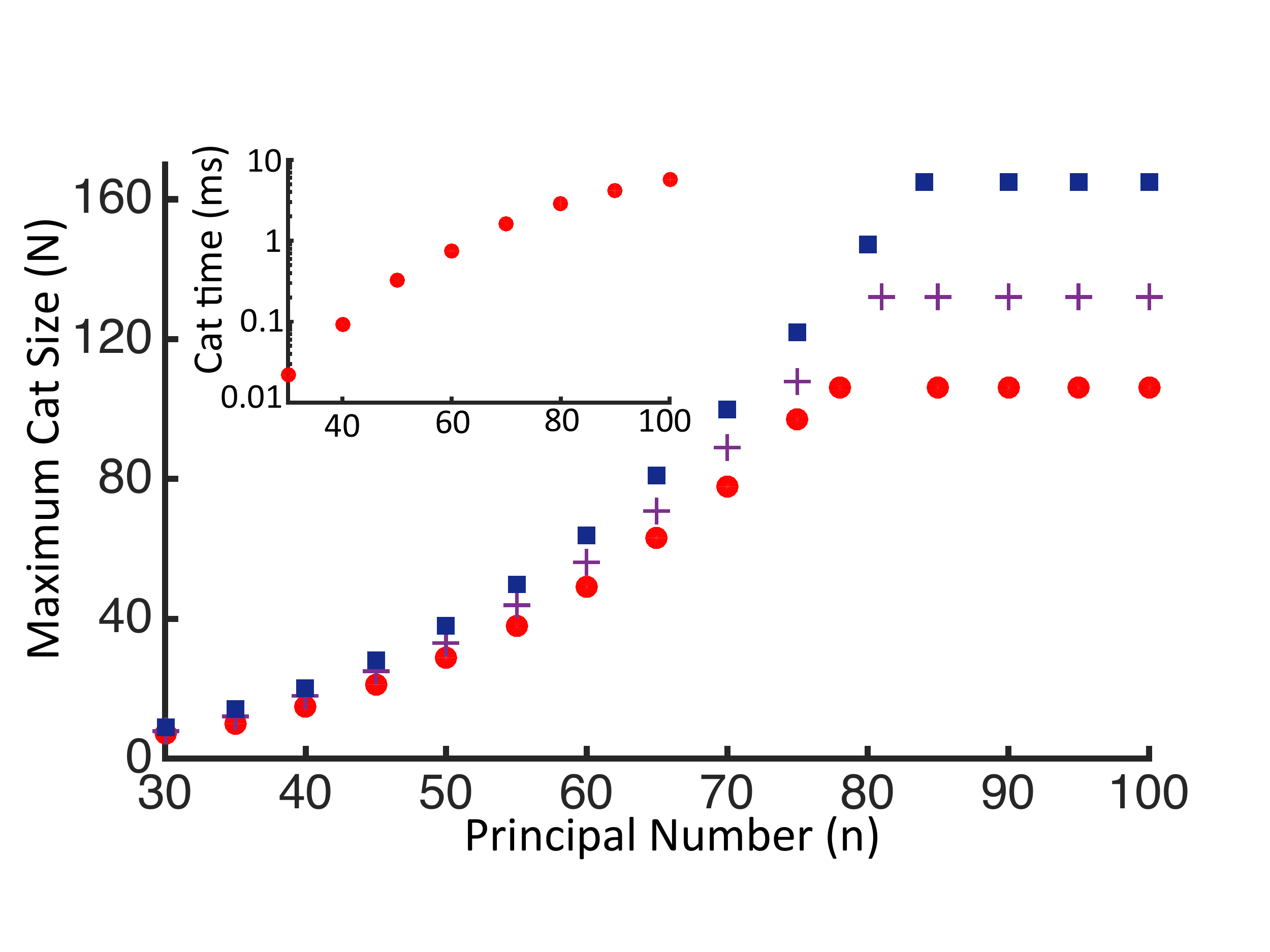}}
\caption{ (Color online) Maximum achievable cat size as a function of the principal number $n$ of the Rydberg state. Rydberg state decay is adjusted to cause 20\% infidelity. The interaction inhomogeneity is set to create less than 1\% infidelity, see Fig. 3, and the higher-order nonlinearities are set to create 10\% (red circle), 20\% (purple plus) and 30\% (blue square) infidelity, see Fig. 2. The inset shows the required cat creation time as a function of $n$ for the case where the higher-order nonlinearities cause 10\% infidelity.} \label{Nvsn}
\end{figure}

Taking into account the mentioned imperfections, Fig.~\ref{Nvsn}  shows the achievable cat size as a function of the principal number $n$. Up to $n \sim 80$, the size increases with $n$.
Higher $n$ leads to a stronger interaction, hence allowing weaker dressing, and to smaller loss, favoring the creation of larger cats. However, for $n \sim 80$ the diminishing spacing between neighboring Rydberg levels (which scales like $n^{-3}$) limits the detuning and hence the interaction strength, since $\chi_0=2w^4 \Delta$ and $w$ has to be kept small, see Fig.~\ref{wVsN}. As a consequence,
larger cat states cannot be achieved at higher principal numbers.

Here we justify the behavior of Fig.~\ref{Nvsn} in a more detailed scaling argument. For a constant fidelity the maximum achievable cat size $N$ at each principal number $n$ is limited by Rydberg decay, $F_{dc}=e^{-\lambda }$ where $\lambda = 0.95 \frac{N}{2} \tau_{c} \gamma_{\tilde{e}} $. Let us analyze how $\lambda$ scales with $N$ and $n$.
The Rydberg decay rate scales as
$\gamma_{|\tilde{e}\rangle} \varpropto w^{2} n^{-3}$.
In order to have a constant non-linearity fidelity of $F_{nl}=0.8$, the dressing strength $w$ has to scale like $N^{-0.84}$, see Fig.~\ref{wVsN}.
The cat creation time $\tau_c=\frac{\pi}{\chi_0} \varpropto w^{-4}\Delta^{-1}$ scales differently before and after the transition point $n \sim 80$. Before the transition point the scaling of $\Delta$ can be obtained by noting that the trap size is a fraction of the blockade radius, $\Delta=\frac{C_{6}}{2R_{b}^{6}} \varpropto \frac{n^{11}}{N^{2}}$, where the exact value of the fraction coefficient is determined by $F_{IH}$, see Fig.~\ref{Inhomogeneity Fig}. Therefore we conclude that $\lambda \varpropto \frac{N^{4.7}}{n^{14}}$, which states that before the transition point larger cat states are realizable by dressing to higher principal numbers, $N \varpropto n^{3}$ for constant fidelity. However, after the transition point the small level spacing imposes a limit on the detuning, $\Delta \varpropto n^{-3}$. Therefore after the transition point $\lambda \varpropto N^{2.7}$, which is independent of $n$. This prevents the realization of larger cat states at higher principal numbers.

One sees that superposition states of over 100 atoms are achievable with good fidelity. In Fig.~4 the interaction inhomogeneity is tuned to create less than $1\%$ infidelity. Dressing to an $S$ orbital is desired due to its isotropic interaction in the presence of trap fields. In Fig.~\ref{Nvsn}, after the transition point in $n$ the detuning is chosen such that 90\% of the Rydberg component of the dressed state is $5sns \,^{3}S_{1}$. Note that without a cryogenic environment the maximum achievable cat size in Fig.~\ref{Nvsn} would be reduced from 165 to 120 atoms, see Sec.~\ref{BBR}.

\section{Experimental Realization}
\label{Realization}
Experimental implementation of our scheme seems feasible. Rydberg excitations in Strontium have been realized over a wide range up to $n=500$
\cite{Sr-Rydberg 1,Sr-Rydberg 2,Sr-Rydberg 3,Sr-Rydberg 4}.
Rydberg dressing of two atoms has been used to create Bell-state entangled atoms \cite{Dressing-realization1}. Recently Rydberg dressing of up to 200 atoms in an optical-lattice has been reported \cite{Dressing-realization2}, where the collective interaction was probed using interferometric techniques. Ref. \cite{Dressing-realization2} also identified a collective many-body decay process, which is however not a limiting factor for our scheme, as discussed in Sec.~\ref{collective decoherence}. 

The Rydberg state $5sns \, ^{3}S_{1}$ is accessible from the $5s5p \, ^{3}P_{0}$ level with a $317$nm laser field. The required Rydberg transition Rabi frequency $\Omega_r/2\pi$ (up to 15 MHz) can be obtained with a tunable single-frequency solid state laser of Ref. \cite{Dressing-laser}. The relatively large detuning values ($ 4$MHz$<\Delta/2\pi<340 $MHz in Fig. 4) make the interaction stable against Doppler shifts.


Fulfilling the adiabaticity condition discussed in Sec.~\ref{Ker-type} is not difficult.
In a highly adiabatic example, $\frac{\dot{\theta}}{E_+}=0.01$, the dressing laser can be switched from zero to $\frac{\Omega_r}{2\pi}= 15$ MHz over 18 ns (for $\frac{\Delta}{2\pi}=270$ MHz and 165 atoms). For this example,  99.991\% of the population returns to the ground state at the end of dressing, so adiabaticity is almost perfect. This adiabatic switching time of 18 ns is many orders of magnitude shorter than the related cat creation time of 1.4 ms.
Adequate interaction timing precision is also required to align the created cat on the equator of Bloch sphere as explained in Sec.~\ref{Steps Towards the Energy Cat}.
 For the 165-atom cat state mentioned above, a timing precision of order $\delta \tau_c=\frac{2w^2\,\delta\phi}{\chi_0}=\frac{4\Delta}{5\sqrt{N}\Omega_r ^2} \approx 7.5$ ns is required  for a  phase precision of order $\delta\phi=\frac{1}{5\sqrt{N}}=\pi/150$.

The Husimi Q function can be reconstructed based on tomography, i.e. counting atomic populations after appropriate rotations on the Bloch sphere. Modern fluorescence methods can count atom numbers in the required range with single-atom accuracy \cite{Dressing-realization2,detection}.

\section{Other sources of imperfection}
\label{More imperfection}

\subsection{Effects of Atomic Motion in the Optical Lattice}
\label{atomic motion}

Laser manipulation of the atomic state leads to phases that depend on the atomic position. Atomic motion could therefore lead to decoherence. To suppress this effect, in the present proposal the atoms are confined to the ground states of the optical lattice traps. As a consequence, all position-dependent phase factors are constant over the course of the experiment and can be absorbed into the definition of the excited states. We now explain these points in more detail. Let us consider the $j^{th}$ atom, and let us assume that it is initially in the ground state (zero-phonon state) of its optical lattice site. We will denote the corresponding state $|g\rangle_j  |0\rangle_j$. Applying the part of the Hamiltonian that is due to the laser to this
state gives $(\Omega_e(t)\ e^{ik\hat{x}_j}|e\rangle_j\langle g|)|g\rangle_j|0\rangle=\Omega_e(t)|e\rangle_je^{ik\hat{x}_j}|0\rangle_j$.
We can rewrite
the position operator $\hat{x}_j$ as the sum of the constant position of the $j^{th}$ site of the
trap $(x_{0j})$ plus a relative position operator $\hat{\xi}_j=s(\hat{a}_j^{\dagger}+\hat{a}_j)$, where $s=\sqrt{\frac{\hbar}{2m\omega_{tr}}}$ is the spread of the ground state wave function, $\omega_{tr}$ is the trap frequency and $(\hat{a}_j,\hat{a} _j^{\dagger})$ are the phononic annihilation-creation operators of the $j$th atom. In the Lamb-Dicke regime $(\eta=\frac{ks}{\sqrt{2}}\ll1)$  one can expand the exponential to get
\begin{equation}
 e^{ik\hat{x }_j}=e^{ikx_{0j}}e^{ik\hat{\xi}_j}=e^{ikx_{0j}}(l+i\eta(\hat{a}_j+\hat{a}_j^{\dagger} )+O(\eta^2)).
\end{equation}
The phase factor $e^{ikx_{0j}}$ is constant over the course of the experiment and can be absorbed into the definition of the atomic basis states by defining $|e'\rangle_j \equiv e^{ikx_{0j}}|e\rangle_j$. The Hamiltonian describing the laser excitation can now be written
in the new basis $|g,0\rangle_j, \, |e',0\rangle_j,\, |e',1\rangle_j$ as:
\begin{equation}
\left(\begin{array}{ccc}
0 & \Omega_{e} & \eta\Omega_{e}\\
\Omega_{e} & 0 & 0\\
\eta\Omega_{e} & 0 & \omega_{tr}
\end{array}\right)\left(\begin{array}{c}
|g,0\rangle_j\\
 |e',0\rangle_j\\
 |e',1\rangle_j
\end{array}\right)
\end{equation}
Starting from the spin and motional ground state $|g,0\rangle_j$, the probability of
populating the state $ |e',1\rangle_j$, corresponding to the creation of a phonon, will be negligible if $\Omega_e \eta \ll \omega_{tr}$. With the parameters that we considered in our proposal ($\Omega_e\sim 1$ kHz, $\eta=0.1, \frac{\omega_{tr}}{2\pi} \sim 400$ kHz)  \cite{3D Lattice} the population of $ |e',1\rangle_j$  will be eight orders of magnitude smaller than the population in the motional ground state.

\subsection{Effects of High Density}
\label{high density}

The relatively small lattice spacing of order $200$nm might raise concerns about molecule formation and level mixing. At high atomic densities there is another potential loss channel, Rydberg molecule formation \cite{Pfau-Molecule}. Molecule formation only occurs when the attractive potential due to Rydberg electron-neutral atom scattering moves the two binding atoms to a very small separation (of order 2nm), where the binding energy of the molecules can ionize the Rydberg electron and form a Sr$^2_{+}$ molecule \cite{Ott-Molecule}. Without the mass transport, stepwise decay or ionization of the Rydberg atom is ruled out by the quantization of Rydberg state, as has been discussed and experimentally tested in \cite{Pfau-Molecule}, because even at high densities the small molecular binding energy of nearby atoms is orders of magnitude smaller than the closest Rydberg levels for all the principal numbers. The occurrence of ion pair formation is also highly unlikely in this system \cite{Ott-Molecule}. We propose that confining the atoms by an optical lattice can prevent the described mass transport and completely close the molecule formation loss channel. High atomic density can also lead to strong level mixing at short distances \cite{level mixing,level mixing2}. However, the experiment of Ref. \cite{level mixing 3} shows that the plateau-type interaction can persist in the presence of strong level mixing because most molecular resonances are only weakly coupled to the Rydberg excitation laser.

\subsection{Effects of Blackbody Radiation}
\label{BBR}

\begin{figure}
\scalebox{0.34}{\includegraphics*[viewport=1 100 800 480]{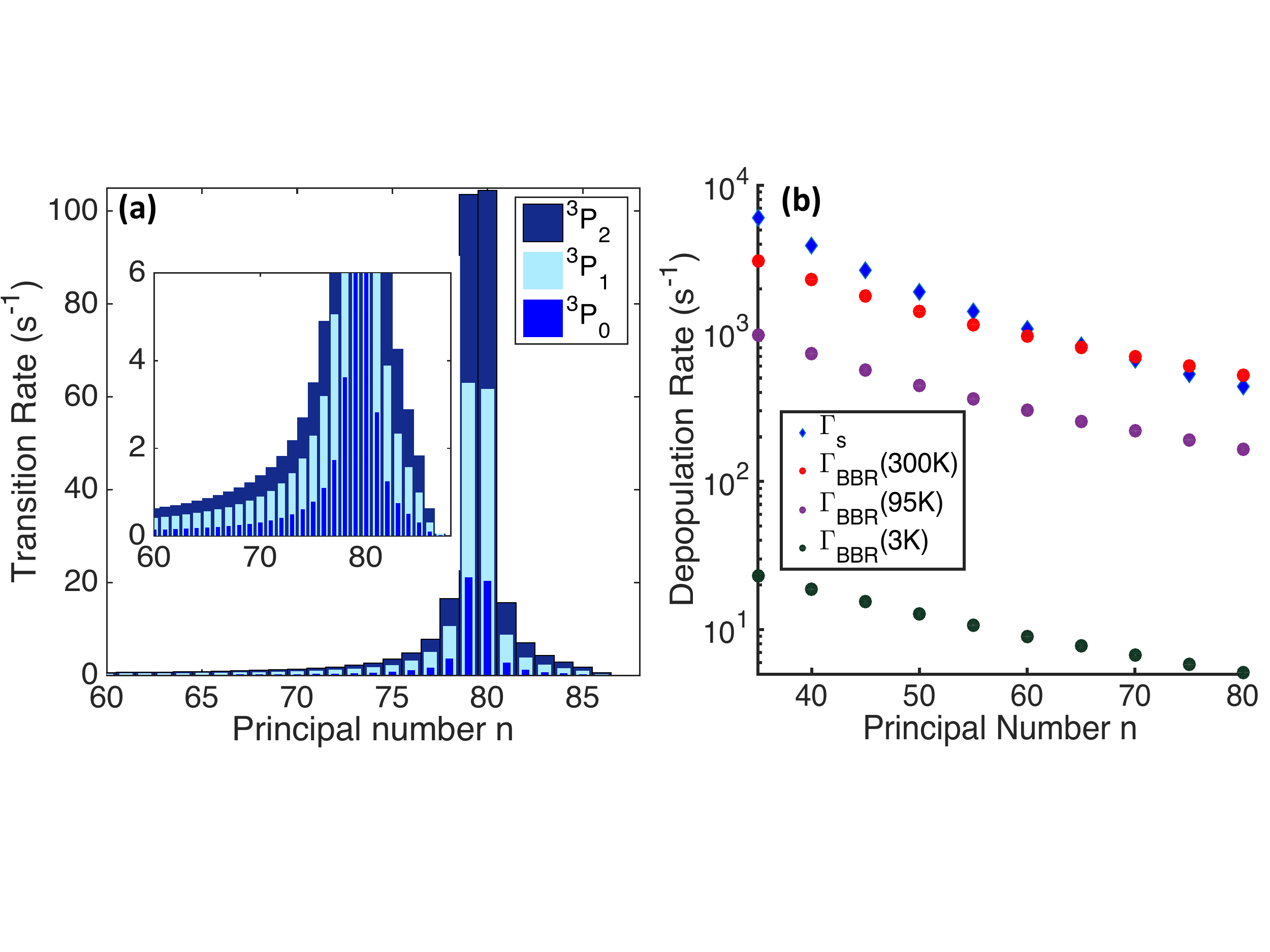}}
\caption{ (Color online)  Depopulation of Strontium Rydberg levels due to blackbody radiation (BBR) induced transitions. a) BBR-induced transition rates (Einstein-B coefficients) from $5s80s\,^3S_1$ to the neighboring $5snp\, ^3P_2$ (dark blue), $5snp\, ^3P_1$ (light blue), $5snp\, ^3P_0$ (blue) levels. The sum of these transition rates gives the total BBR-induced depopulation rate $\Gamma_{BBR}$. The inset is a 20 times enlarged view. b) Rydberg depopulation rates due to spontaneous decay ($\Gamma_s$ shown in blue diamond) and BBR-induced transitions ($\Gamma_{BBR}$) at environment temperatures of 300K (red circle), 95K \citep{Cryogenic1} (purple circle), and 3K \citep{Cryogenic2} (green circle) as a function of the principal number. The use of a cryogenic environment significantly suppresses the unwanted effects of BBR.} \label{BBR-figure}
\end{figure}

Blackbody radiation (BBR) could reduce the lifetime by transferring the Rydberg state population to neighboring Rydberg levels (with both higher and lower principal numbers $n$) as illustrated in Fig.~\ref{BBR-figure}a. The BBR-induced transition probability is given by the Einstein B-coefficient $\Gamma_{BBR}=\sum\limits_{f}  B_{if}=\sum\limits_{f} \frac{A_{if}}{e^{\frac{\hbar \omega_{if}}{k_B T}}-1}$ \cite{A-coefficient1,A-coefficient2,A-coefficient3}, where $T$ is the environment temperature, $k_B$ is the Boltzmann constant and both $\omega_{if}$ and $A_{if}$ are defined in Sec.~\ref{A coefficient}.

At the environment temperatures of 300K, 95K \citep{Cryogenic1} and 3K \citep{Cryogenic2}, including the BBR-induced transitions increases the total decoherence rate $\Gamma_{\tilde{e}}$ by 120\%, 40\% and 1\% (see Fig.~\ref{BBR-figure}b) for $n \approx 80$, which results in maximum achievable cat sizes of 120, 150 and 165 atoms respectively (considering $F_{nl}=0.7, \, F_{dc}=0.8$).
Note that cryogenic environments with 95K and  1K were used in a Strontium lattice clock experiment \citep{Cryogenic1} and in a cavity QED experiment with Rydberg atoms \cite{Cryogenic2} respectively.

BBR could also disturb the Ramsey-type interferometry used for detecting energy decoherence by producing an AC stark shift; this effect is  quantified in section \ref{Energy-decoherence}. Furthermore, BBR-induced decoherence could be inhomogeneous due to temperature inhomogeneities in the environment. This would introduce unwanted coupling to non-symmetric Dicke states in the cat creation process. The use of a cryogenic environment  significantly suppresses these effects as well.

\subsection{Effects of Collective Many-body Decoherence}
\label{collective decoherence}

BBR-induced transitions to neighboring Rydberg levels (see Fig.~\ref{BBR-figure}a) can also lead to collective many-body decoherence \cite{LineBroadening, Dressing-realization2}. The interaction between the target $nS$ Rydberg level and some of the populated neighboring $n'P$ levels is of a strong long-range dipole-dipole type due to the formation of F\"{o}rster resonances. This strong interaction causes an anomalous broadening \cite{LineBroadening}. The mentioned decoherence process only starts after the first BBR-induced transition occurs.
However, the weak dressing strength and small ensemble size $(N<200)$ in our scheme make the probability of populating the target Rydberg state and consequently neighboring Rydberg levels very small. For example at the environment temperatures of 300K, 95K and 3K and for dressing to $n\approx 80$, the probabilities of not populating the strongly interacting neighboring Rydberg levels over the cat creation time for cat sizes of 120, 150 and 165 atoms respectively are $P_{BBR}(0)=exp(-\frac{N}{2}w^2 \Gamma_{BBR} \tau_c)=$98.63\%, 99.26\% and 99.96\% respectively. It has been observed in the realization of many particle Rydberg dressing \cite{Dressing-realization2} that when the transition probability is low enough (of the order of $P_{BBR}(0)\geq 82\%$, as can be calculated from the information provided in Ref. \cite{Dressing-realization2}) the many-body decoherence effects are negligible and decoherence rate is dominated by the Rydberg depopulation rate (see Sec.~\ref{Decoherence}).

\section{Testing Energy Decoherence}
\label{Energy-decoherence}
In the context of modifications of quantum physics, decoherence in the energy basis is quite a natural possibility to consider \cite{Milburn,Gambini,Blencowe}. It is usually introduced as an additional term in the time evolution for the density matrix that is quadratic in the Hamiltonian, $\frac{d\rho}{dt}=\frac{i}{\hbar}[H,\rho]-\frac{\sigma}{\hbar^2}[H,[H,\rho]]$, which leads to a decay of the off-diagonal terms of the density matrix in the energy basis according to $\rho_{nm}(t)=\rho_{nm}(0)e^{-i\omega_{nm}t}e^{-\gamma_E t}$ \cite{Gambini}, where $ \gamma_E=\sigma \omega_{nm}^2$. Here $\omega_{nm}$ is related to the energy difference of the two componants and $\sigma$ can be interpreted as a timescale on which time is effectively discretized, e.g. related to quantum gravity effects. It is of interest to establish experimental bounds on the size of $\sigma$, which could in principle be as small as the Planck time ($10^{-43}$ s). 

The corresponding decoherence rate for the energy cat in this proposal would be $\gamma_{E}=\sigma (\frac{N\Delta E}{\hbar})^{2}$, where $\Delta E$ is the energy difference between the ground and excited state of each qubit, and $N$ is the cat size. To detect the energy decoherence one prepares the energy cat state, followed by a waiting period. To observe the decoherence effect, one detects the Ramsey fringes for the revival. The visibility of the Ramsey interference is also sensitive to other decoherence sources,  where in the absence of dressing laser the dominant ones are the trap loss rate $\Gamma$, which reduces the visibility by a factor $\exp(-N \Gamma t)$, and phase diffusion that is explained below.


The large energy difference of the cat state increases the sensitivity of the Ramsey interferometry that we are using for the detection of energy decoherence. Therefore, it is  important to consider the effect of fluctuations in the detuning between the laser and the atomic transition. Let us first note that the cat state is more sensitive to multi-particle (correlated) than to single-particle (uncorrelated) noise, which  results in a phase diffusion affecting the visibility of Ramsey fringes by $e^{-N^2\delta_c^2t^2}$ and $e^{-N \delta_{uc}^2t^2}$ respectively \cite{phase-diffusion}. Comparing the two cases, correlated fluctuations should be $\sqrt{N}$ times more stabilized than uncorrelated fluctuations. The most important source of noise in our system is the fluctuation of the laser frequency. A probe laser linewidth as narrow as 26 mHz \cite{laser1} has been achieved in optical atomic clock experiments, and there are proposals for much smaller linewidths \cite{laser2,laser3} with recent experimental progress \cite{laser4},
justifying our example of a 10mHz linewidth, see below.

Other sources of multi-particle and single-particle noise have been well studied in the context of Strontium atomic clocks \cite{atomic-clocks-noise1,atomic-clocks-noise2} and are comparatively negligible. Here we address a few of them in our scheme.
One of the noise sources is the trap field’s intensity fluctuation; however, using the magic wavelength makes the atomic transition frequency independent of the trap laser intensity. Considering the variation of the Stark shifts  due to the trap laser as a function of frequency at the magic wavelength \cite{Blue-detuned}, the relative scalar light shifts could be kept within 0.1mHz uncertainty by applying a trap laser with a 1MHz linewidth. In addition to the scalar light shift, the inhomogeneous polarization of trap fields in 3D optical lattices can result in an inhomogeneous tensor light shift \cite{W. Happer}; however, the use of the bosonic isotope $^{88}$Sr with zero magnetic moment cancels the tensor light shift \cite{T. Akatsuka} in our scheme.
Environmental temperature fluctuations ($\delta T$) also lead to atomic frequency fluctuations that are proportional to $T^3 \delta T$ due to the BBR-induced light shift \cite{atomic-clocks-noise1}. This is another reason why a cryogenic environment is advantageous. For example controlling the  environment temperature of $95$K \citep{Cryogenic1} to within a range of $\delta T=1K$ keeps the BBR-induced noise shift below $1$ mHz.

A conservative estimate of the experimentally measurable energy decoherence rate can be obtained by considering the case where the energy decoherence dominates all other decoherence sources during the waiting period. Increasing the cat size $N$ is helpful because it allows one to enhance the relative size of the energy decoherence contribution. For example, choosing $t \propto N^{-1}$ keeps the loss and phase diffusion contributions fixed, while the energy decoherence still increases proportionally to $N$.
Using a cat state with $N=165$ atoms (see Fig. 4), which corresponds to $N \Delta E=$ 300 eV, assuming a laser linewidth of $10$ mHz (see above), and considering a trap loss rate of $\Gamma=10$ mHz \cite{Pfau-Molecule}, the minimum detectable discretization time scale $\sigma$ is of order  $10^{-34}$ s. This would improve the measurement precision by 4 and 11 orders of magnitude compared to what is possible based on Ref. \cite{ion-Cat} and Ref. \cite{Ghobadi} respectively.

\appendix
\section{Effects of Interaction Inhomogeneity}
\label{supp}
Here we explain the steps in calculating the effects of inhomogeneous interaction on the cat state's fidelity $F_{IH}=|\langle \eta=1|e^{-i\hat{V}_{IH}\tau_c}|\eta=1\rangle|^2$ (see Sec.~\ref{sec:Inhomogeneities}). Taylor expanding $e^{-i\hat{V}_{IH}\tau_c}$ and considering the expectation values $\langle \eta=1|\hat{\sigma}^i_{ee}|\eta=1\rangle=1/2$ and $\langle \eta=1|\hat{\sigma}^i_{ee} \hat{\sigma}^j_{ee}|\eta=1\rangle=\frac{1}{4}+\frac{\delta_{ij}}{4}$, one obtains an estimate for the fidelity. The first order of the expansion is zero because we defined  $\epsilon_{ij}$ as fluctuations around a mean value. The second order can be calculated using $\langle \eta=1|\hat{V}_{IH}^2|\eta=1\rangle=\frac{1}{2}\sum\limits_{i \neq j} \frac{1}{2}\sum\limits_{l \neq m} C_{ijlm} \epsilon_{ij}\epsilon_{lm} $, where $C_{ijlm}=1/16$ if all the indices are unequal, $C_{ijlm}=1/8$ if there is a pair of equal indices, and $C_{ijlm}=1/4$ when there are two pairs of equal indices. The convergence of the expansion for the fidelity can be tested numerically.
In Fig. 3b of the paper the ratio of the third order to the second order of the expansion for $F_{IH}$ is $\frac{O(3)}{O(2)}=10^{-6}, 5\times10^{-5}, 8\times10^{-4}, 8\times10^{-3} $ for $\frac{D}{R_b}=0.1, 0.2, 0.3, 0.4$ respectively, suggesting good convergence in this regime.

\section{Dipole Matrix Elements}
\label{dipole matrix elements}
In Strontium one needs to consider both valence electrons $(|i \rangle=|n_{1i}n_{2i}l_{1i}l_{2i}L_{i}S_{i}J_{i}M_{i} \rangle)$ in the calculation of the dipole matrix elements \cite{A-coefficient3}
\begin{eqnarray}\label{dipole}
\begin{array}{c}
|\langle i|\vec{r}|f\rangle|^2= \max(l_{2i},l_{2f})\,(2L_{f}+1)(2J_{f}+1)(2L_{i}+1)
\\
\left\{ \begin{array}{ccc}
J_{f} & 1 & J_{i}\\
L_{i} & S & L_{f}
\end{array}\right\}^2
\left\{ \begin{array}{ccc}
L_{f} & 1 & L_{i}\\
l_{2i} & l_{1i} & l_{2f}
\end{array}\right\}^2
|\langle n_{2i}l_{2i}|r|n_{2f}l_{2f}\rangle|^{2},
\end{array}
\end{eqnarray}
where $L$ and $S$ are the total orbital angular momentum and spin, $l$ and $s$ refer to individual electrons, and $J$ and $M$ refer to total angular momentum. The active electron in the transition is labeled by 2, and $\langle n_{2i}l_{2i}|r|n_{2f}l_{2f}\rangle$ is the radial dipole matrix element between initial and final state, and the curly bracket is a Wigner-6j symbol.

\begin{acknowledgments}
We acknowledge financial support from AITF and NSERC. We thank P.~Grangier, M.~Saffman and B. Sanders for fruitful discussions.
\end{acknowledgments}

\end{document}